\documentclass[iop]{emulateapj}

\shorttitle{Broadband emission spectrum of WASP-19b}
\shortauthors{Zhou et al.}

\usepackage{graphicx}
\usepackage{amsmath}
\usepackage{xspace}

\newcommand{\chisq}{$\chi ^2$\xspace}
\newcommand{\icarus}{Icarus}
\newcommand{\myemail}{george@mso.anu.edu.au}

\begin{document}

\title{Examining the broadband emission spectrum of WASP-19b: A new $z$ band eclipse detection}

\author{George Zhou\altaffilmark{1},
Lucyna Kedziora-Chudczer\altaffilmark{2},
Daniel D.R. Bayliss\altaffilmark{1},
Jeremy Bailey\altaffilmark{2}}

\altaffiltext{1}{Research School of Astronomy and Astrophysics, Australian National University, Cotter Rd, Weston Creek, ACT 2611, Australia; \email{\myemail}}
\altaffiltext{2}{School of Physics, University of New South Wales, Sydney, NSW 2052, Australia}

\begin{abstract}
WASP-19b is one of the most irradiated hot-Jupiters known. Its secondary eclipse is the deepest of all transiting planets, and has been measured in multiple optical and infrared bands.  We obtained a $z$ band eclipse observation, with measured depth of $0.080\pm0.029\,\text{\%}$, using the 2\,m Faulkes Telescope South, that is consistent with the results of previous observations. We combine our measurement of the $z$ band eclipse with previous observations to explore atmosphere models of WASP-19b that are consistent with the its broadband spectrum. We use the VSTAR radiative transfer code to examine the effect of varying pressure--temperature profiles and C/O abundance ratios on the emission spectrum of the planet. We find models with super-solar carbon enrichment best match the observations, consistent with previous model retrieval studies. We also include upper atmosphere haze as another dimension in the interpretation of exoplanet emission spectra, and find that particles $<0.5\mu\text{m}$ in size are unlikely to be present in WASP-19b. 
\end{abstract}

\keywords{planets and satellites: atmospheres -- planets and satellites: individual (WASP-19b)}

\section{Introduction}
\label{sec:introduction}

Recent observations of transiting planet systems have lead to the first in-depth characterisation of exoplanet atmospheres. Observations of the secondary eclipse event, when the planet is blocked by the host star, is the predominant method of measuring the emergent flux of close-in exoplanets. In particular, secondary eclipses observed at multiple wavelength bands have provided first spectral energy distribution of exoplanets \citep{2005ApJ...626..523C,2008ApJ...686.1341C}. Similar observations have revealed the presence of molecular absorption features \citep[e.g.][]{2008Natur.456..767G,2009ApJ...690L.114S} in the emission spectra of hot-Jupiters, and hinted at the diversity of chemical compositions across exoplanet atmospheres \citep[e.g.][]{2008ApJ...676L..61B,2009ApJ...707...24M}.

WASP-19b \citep{2010ApJ...708..224H} is a $1.17\,M_\text{Jup}$, $1.39\,R_\text{Jup}$ exoplanet in a 0.79 day prograde orbit \citep{2011ApJ...730L..31H} which transits a $V_\text{mag}=12.3$ G dwarf. The equilibrium temperature for the planet is at least 2000\,K, making it one of the hottest hot-Jupiters known, and the most favourable target for eclipse observations. The proximity of WASP-19b to the host star also makes it an interesting case study of irradiated atmospheres. In particular, eclipse observations have shown that WASP-19b is inconsistent with the hypothesis that highly irradiated planets exhibit thermal inversion features \citep{2003ApJ...594.1011H,2007ApJ...668L.171B,2008ApJ...678.1419F}, although some exceptions are known  \citep[e.g. HD189733b, TrES-3b, and XO-1b,][]{2008ApJ...686.1341C,2010ApJ...711..374F,2008ApJ...684.1427M}. \citet{2012ApJ...758...36M} proposed WASP-19b as a planet hosting a carbon rich atmosphere, depleted in TiO, which is a primary absorber for inversion layers. In addition, the Carbon-to-Oxygen ratio (C/O) is a potential indicator for the location in the proto-planetary disk where these hot-Jupiters originated \citep[e.g.][]{2004ApJ...611..587L,2011ApJ...743L..16O}. 

The C/O enrichment hypothesis is based upon existing multi-band eclipse observations of WASP-19b, including the ASTEP400 broadband centred at $0.67\,\mu\text{m}$ \citep{2013arXiv1303.0973A}, $z$ band \citep{2012ApJS..201...36B,2012arXiv1212.3553L}, $1.190\, \mu \text{m}$ narrow band \citep{2012arXiv1212.3553L}, $H$ band \citep{2010A&amp;A...513L...3A}, $K$ band \citep{2010MNRAS.404L.114G}, Spitzer 3.6, 4.5, 5.8, and 8.0\,$\mu\text{m}$ bands \citep{2011arXiv1112.5145A}, as well as spectrophotometric observations at $1.25-2.35\,\mu\text{m}$ by \citet{2013arXiv1303.1094B}. However, it is difficult to produce a single model that can fit all the measurements within their uncertainty constraints. Ground based observations at the 0.1\% level remain difficult, and are affected by a range of systematic effects, such as atmospheric variations, unstable telescope tracking, and detector defects. Independent confirmation observations are required to strengthen the reliability of individual measurements.

The depth of the $z$ band eclipse is particularly important in determining the C/O ratio of WASP-19b, a deeper eclipse is indicative of an atmosphere deficient in TiO absorption and enriched in C/O abundance. New Technology Telescope ULTRACAM observations by \citet{2012ApJS..201...36B} reported an eclipse depth of $0.088\pm0.019\,\text{\%}$, whilst a combined set of observations with EulerCam and TRAPPIST over ten epochs by \citet{2012arXiv1212.3553L} reported a shallower eclipse depth of $0.035\pm0.012\,\text{\%}$. Whilst these observations are consistent at the $\sim2\sigma$ regime, the difference between the two measurements makes it difficult to constrain the atmosphere models of WASP-19b.

In this study, we present an independent observation of a WASP-19b eclipse event using Faulkes Telescope South aimed at confirming its $z$ band secondary eclipse depth. We present a careful treatment of the photometry to achieve near photon-limited lightcurves. To investigate the previous claim of a carbon rich atmosphere, we use the VSTAR radiative transfer code \citep{2012MNRAS.419.1913B} and the ensemble of observations to model the atmosphere of WASP-19b and examine the effects of temperature-pressure profiles and C/O abundance on its emergent spectrum. One draw back of existing model retrieval studies \citep{2009ApJ...707...24M} is its lack of treatment for non-isotropic scattering. The lack of absorption features in the transmission spectrum in HD189733b \citep{2008MNRAS.385..109P}, as well as the weaker than expected detections of sodium in various hot-Jupiters \citep[e.g.][]{2002ApJ...568..377C,2003ApJ...589..615F,2012MNRAS.426.2483Z}, all point to the importance of clouds and haze in modelling planetary atmospheres. We also exploit the rigorous treatment of Rayleigh scattering by VSTAR to investigate the effect of upper atmosphere haze on the emission spectrum of WASP-19b.

\section{Detection of $z$ band eclipse}
\label{sec:zband}

\subsection{Observations}
\label{sec:observations}
\begin{figure}[h]
  \centering
  \includegraphics[width=8cm]{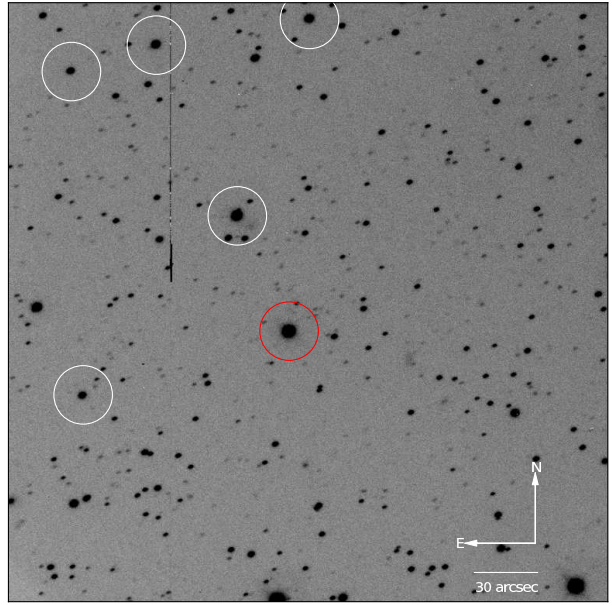}
  \caption{FTS $z$ band image of the WASP-19 field. The target, located in the centre of the field, is circled in red; the chosen set of reference stars are circled in white. The size of the circle indicates the size of the background aperture. The column of dead pixels in the top left of the image was masked out for the photometry.}
  \label{fig:image}
\end{figure}

We monitored an eclipse of WASP-19b using the 2\,m Faulkes Telescope South (FTS), located at Siding Spring Observatory, Australia, on 2012 December 29, from 12:03--15:50 UT, with the expected eclipse occurring during 13:25--15:02 UT. Observations were performed in the Pan-STARRS $z$-band, centred at $0.866\,\mu\text{m}$ \citep{2012ApJ...750...99T}, using the Merope $2\text{K} \times 2\text{K}$ camera, with $4.7' \times 4.7'$ field of view, unbinned pixel size of $0.139\text{"\, pixel}^{-1}$, read out with $2\times 2$ bins. $161\times60\,\text{s}$ exposures were taken. The seeing on the night previous to the observing sequence was $\sim 1\text{''}$. The telescope was slightly defocused to avoid saturation and to reduce the effect of intra- and residual inter-pixel variations, resulting in point spread functions (PSF) with full width half maximum (FWHM) of $\sim 2\text{"}$. Bias subtraction and flat field corrections were performed using the CCDPROC package in \emph{IRAF}\footnote{IRAF is distributed by the National Optical Astronomy Observatories, which are operated by the Association of Universities for Research in Astronomy, Inc., under cooperative agreement with the National Science Foundation} with the most recent archival calibration frames. An example image is shown in Figure~\ref{fig:image}. 

\subsection{Analysis}
\label{sec:analysis}

\subsubsection{PSF Variations and Adaptive Aperture Photometry}
\label{sec:photometry}

\begin{figure}[h]
  \centering
  \includegraphics[width=8cm]{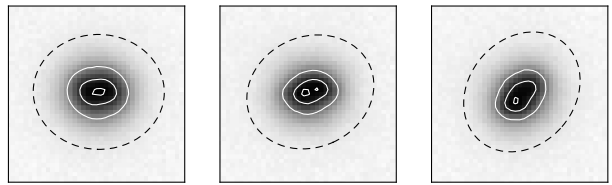}
  \caption{PSF variations of the target star over course of the observations, shown at the start (left), middle (centre), and end of the night (right). The solid white lines mark the 0.95, 0.50, and 0.10 peak height contours. The dashed black ellipse marks the photometry aperture used. The crops are 20 pixels in size.}
  \label{fig:psf}
\end{figure}

Upon close examination of the images, we find the stellar PSF is asymmetric across the image. The distortion and elongation of the PSF is a result of the defocusing applied. In addition, the position angle of the elongated PSF changes with the rotation of the telescope (Figure~\ref{fig:psf}), as FTS is on an alt-az mount. To further investigate the PSF variations, we create a template PSF from a single exposure taken mid-run, and fit it to the remaining exposures, allowing for rotation and spatial dilation. We find no significant deviations in the fit residuals, with the exception of the initial images taken at high airmass, suggesting that the general shape of the PSF remained constant throughout the night.

\begin{figure*}[h]
  \centering
  \includegraphics[width=12cm]{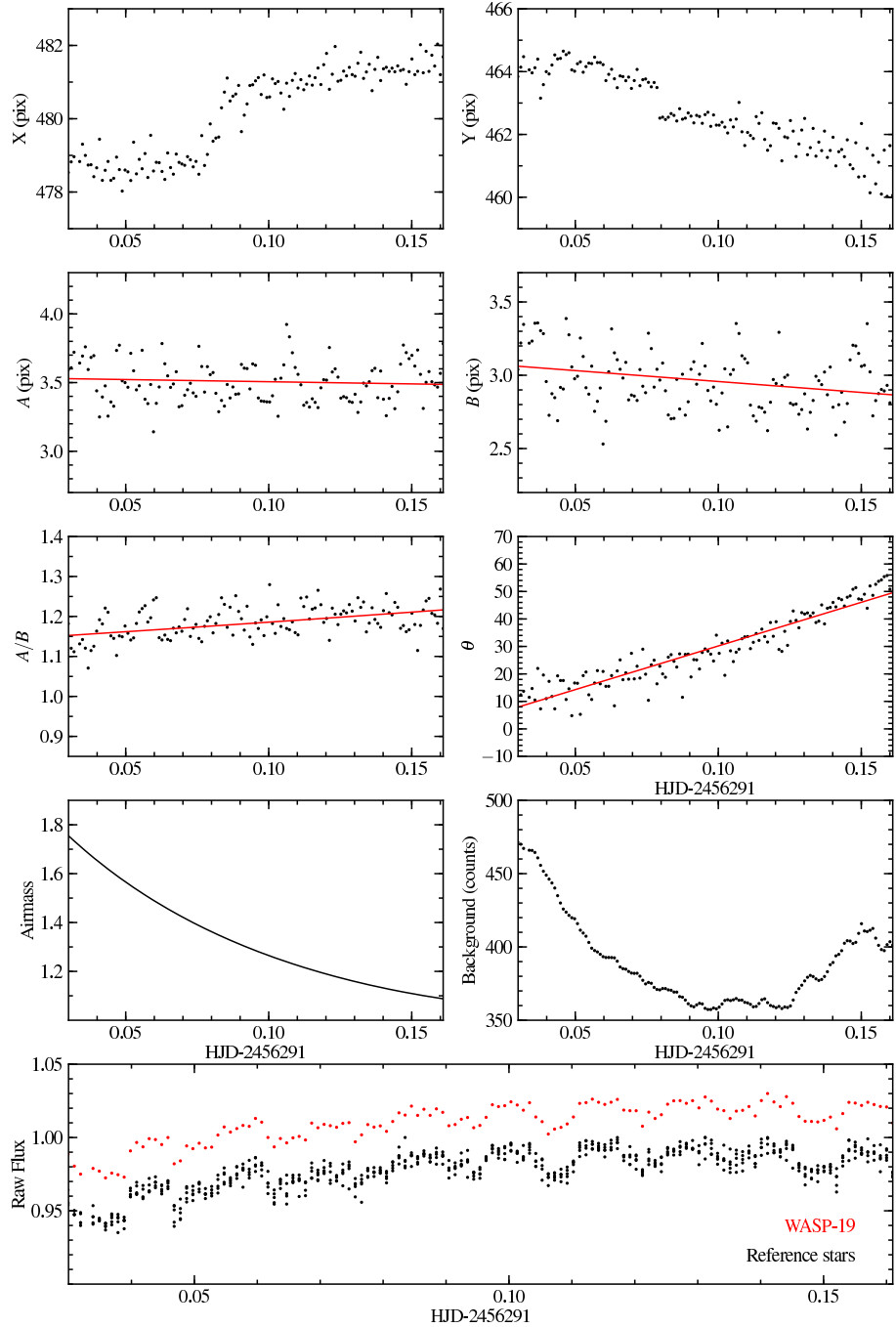}
  \caption{Variations in the target X, Y position, PSF semi-major and minor axis ($A$, $B$), ellipticity $(A/B)$, ellipse position angle $(\theta)$, airmass, background counts, and normalised raw target (red) and ensemble reference (black, arbitrarily offset by 0.05) fluxes are plotted. Linear fits to the ellipse parameters, used to define the elliptical photometry apertures, are plotted in red.}
  \label{fig:params}
\end{figure*}

We performed elliptical aperture photometry on the reduced images. Compared to conventional circular apertures, variable elliptical apertures best account for all of the stellar flux whilst minimising background noise. The ellipse parameters, semi-major axis $A$, semi-minor axis $B$, and position angle $\theta$, were measured using Source Extractor \citep{1996A&amp;AS..117..393B}, and are plotted in Figure~\ref{fig:params}, along with other relevant global parameters. $A$ and $B$ are the maximum and minimum root-mean-square (RMS) of the spatial profile. The size of the aperture, $R$, is a scaling factor that maintains the shape and orientation of the ellipse, and is related to the ellipse parameters $CXX$, $CYY$, and $CXY$ by,
\begin{align}
  \label{eq:radius}
  \begin{split}R^2 =& CXX(x-\bar{x})^2 + CYY(y-\bar{y})^2\\ &+ CXY(x-\bar{x})(y-\bar{y})\end{split}\\
  CXX =& \frac{\cos ^2 \theta}{A^2} + \frac{\sin ^2 \theta}{B^2} \nonumber\\
  CYY =& \frac{\sin^2 \theta}{A^2} + \frac{\cos^2 \theta}{B^2} \nonumber\\
  CXY =& 2\cos \theta \sin \theta \left( \frac{1}{A^2} - \frac{1}{B^2} \right) \, . \nonumber
\end{align}
The lowest out-of-eclipse scatter was achieved using aperture sizes of $R=4.2$, enclosing $\sim 99\text{\%}$ of the flux. The adopted elliptical aperture parameters $A,B,\theta$ were determined from linear fits in time to the averaged measurements from Sourced Extractor. Higher order fits to the ellipse parameters were tested, and did not result in significantly different lightcurves or eclipse depths. Exposures with $\text{HJD} < 2456291.03$ were discarded from the analysis, since they were taken at high airmass, when the PSF shape varied rapidly. 

The background was estimated using a 100 pixel diameter outer aperture on a background image with all detected sources masked out. Since WASP-19 resides in a relatively crowded field, masking out field stars is essential to achieving optimal background subtraction. We note that \citet{2012ApJS..201...36B} followed a similar technique in their analysis. The background count around WASP-19 is plotted in Figure~\ref{fig:params}

Differential photometry was performed using five reference stars (labelled in Fig.~\ref{fig:image}), chosen for their lack of nearby neighbours, similar colour indices to the target, and the eventual stability of the lightcurves. A master reference lightcurve $(M)$ was created by averaging the ensemble of reference stars $(R_i)$, each with errors $\Delta R_i$:
\begin{equation}
  \label{eq:ref_lc}
  M = \sum_i \frac{c_i}{\Delta R_i} R_i \, ,
\end{equation}
where weights $c_i$ were chosen to minimise the RMS scatter of the corrected object lightcurve. The use of weights to minimise the object lightcurve scatter is similar to applying the Trend Filtering Algorithm to the out-of-transit dataset \citep{2005MNRAS.356..557K}. To remove uncorrelated trends in the individual reference star lightcurves, we divided each reference star by a master reference lightcurve made of all other reference stars. Any slow varying residual trends in that reference star were then corrected for by a linear fit. In addition, individual outlier points significantly different from other reference stars were removed by sigma clipping. Finally, a linear trend was removed from the target lightcurve by fitting for the out-of-eclipse points. We note that the target lightcurve was treated by the same processes as the reference lightcurves. The ensemble of raw reference lightcurves, as well as the raw target lightcurve, are plotted in Figure~\ref{fig:params}.

\subsubsection{Eclipse model fitting}
\label{sec:model-fitting}

\begin{figure}[h]
  \centering
  \includegraphics[width=8cm]{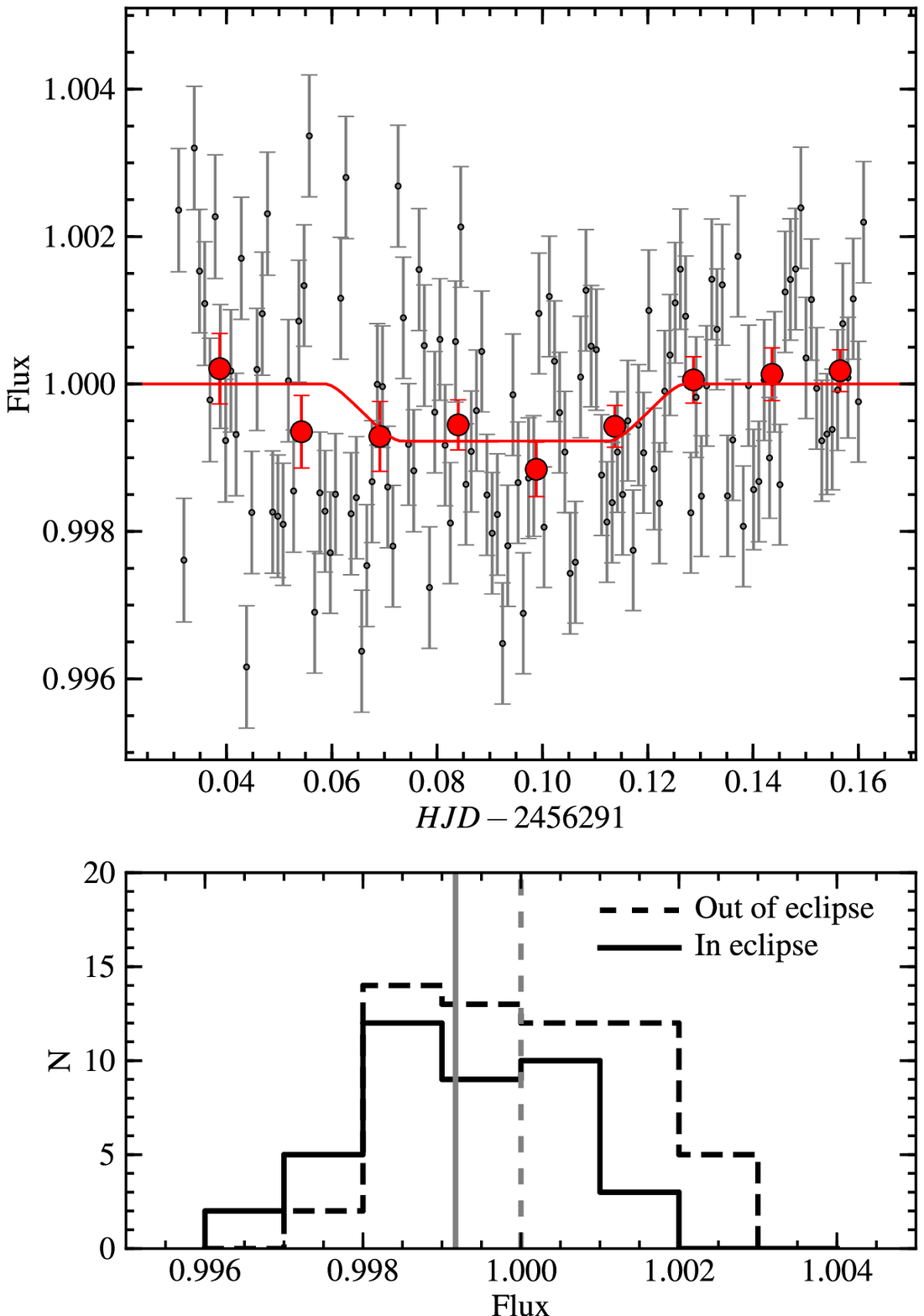}
  \caption{Top: The eclipse lightcurve of WASP-19b, with the best fitting model plotted in red. Data binned at 0.015 days are plotted as large red points for clarity. Bottom: Histogram showing the distribution of flux measurements in- (solid) and out-of-eclipse (dashed), with the centroids of the distributions marked by the corresponding vertical lines.}
  \label{fig:lc}
\end{figure}

We fit a \citet{2002ApJ...580L.171M} eclipse model to the FTS lightcurve via a downhill simplex minimisation of the \chisq of fit, followed by a Markov chain Monte Carlo (MCMC) ensemble sampler \citep[\emph{emcee} implementation,][]{2012arXiv1202.3665F} to determine the uncertainties. For the MCMC routine, we artificially inflate the photometric uncertainties such that the reduced $\chi^2=1$. This accounts for the contribution of other systematic effects in addition to the photon-noise uncertainty. The free parameters of the fit are the transit centre $t_0$, depth $D$, normalised planet orbital radius $a/R_\star$, and the impact parameter $b$. The parameters $t_0,a/R_\star,b$ are constrained by Gaussian priors based on the joint analysis performed by \citet{2011arXiv1112.5145A}. The system period and planet-star radius ratio are fixed to \citet{2011arXiv1112.5145A} values. The fitted lightcurve is shown in Figure~\ref{fig:lc}. 

The final eclipse depth is $0.080\pm0.029\,\text{\%}$.  The corresponding $z$ band brightness temperature  is $2680^{+140}_{-180}\,\text{K}$. A MARCS model atmosphere spectrum \citep{2008A&amp;A...486..951G} was adopted for the host star in the brightness temperature calculation for the planet. The derived brightness temperature agrees well with the ASTEP, 1.6 and 2.09\,$\mu\text{m}$ temperatures \citep{2013arXiv1303.0973A,2011arXiv1112.5145A}. 

The depth can also be derived separately by binning the in- and out-of-eclipse points. In Figure~\ref{fig:lc}, we bin the points according to the predicted ephemeris. The eclipse depth, given by the difference in the sigma clipped mean of the two bins, is $0.083\pm0.026\,\text{\%}$, with the uncertainty taken as the error in the mean of the two bins, added in quadrature. This agrees with the transit depth measured by the model fit.

\subsubsection{Correlation to external parameters}
\label{sec:decorrelation}

Most high-precision transit and eclipse photometry to date have been processed with some form of external parameter decorrelation to remove residual systematic trends. This is often done by multiplying the lightcurve with a linear combination of external parameters, such as airmass, position, FWHM \citep[e.g.][]{2010ApJ...716L..36L,2012ApJS..201...36B}; occasionally, higher order terms have also been employed \citep[e.g. up to 4$^\text{th}$ order][]{2012arXiv1212.3553L}. 

We test for the effectiveness of detrending by simultaneously fitting for the eclipse and a combination of external terms involving X, Y position, semi-major axis $A$, ellipticity $A/B$, airmass, and background counts, whilst holding $t_0$ constant. In each case, the removal of a linear trend is also allowed. Analysis is performed over the entire lightcurve, since the out-of-eclipse points constitute less than half of the observations, and cannot sufficiently represent the entire dataset. Table~\ref{tab:decor-chisq} shows the reduced $\chi^2$ after each minimisation routine. No significant improvements to the \chisq was achieved from any decorrelations. The transit depth also remained roughly independent of these external parameters.

\begin{deluxetable}{lcc}
\tablewidth{0pc}
\tabletypesize{\scriptsize}
\tablecaption{
    Reduced \chisq and eclipse depth after decorrelation
    \label{tab:decor-chisq}
}

\tablehead{
    \multicolumn{1}{c}{External Parameter}          &
    \multicolumn{1}{c}{Reduced $\chi^2$}             &
    \multicolumn{1}{c}{$D$\,\%}      \\
}
\startdata

    None & 3.28 & 0.080\\
    X, Y & 3.25 & 0.087\\
    $A$ & 3.30 & 0.078\\
    $A/B$ & 3.29 & 0.078\\
    $\theta$ & 3.28 & 0.074\\
    Airmass & 3.31 & 0.080\\
    Background & 3.30 & 0.071\\

\enddata 

\end{deluxetable}

\begin{figure}[h]
  \centering
  \includegraphics[width=8cm]{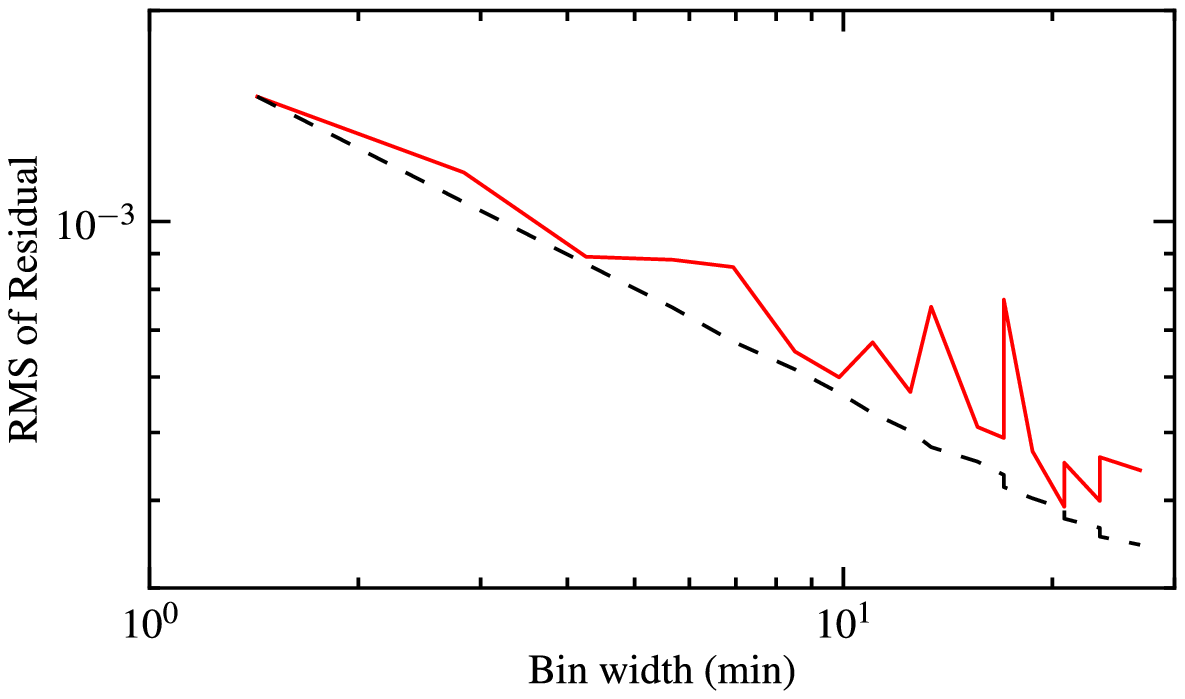}
  \caption{The RMS of the residuals as a function of bin width are plotted in red. The dashed line shows the $1/\sqrt{N}$ drop off expected for an uncorrelated signal.}
  \label{fig:residual_bin}
\end{figure}

We can also check for time correlated noise in the residuals using the $\beta$ factor diagnostic \citep{2008ApJ...683.1076W}. For residuals binned into $M$ bins, with $N$ points per bin, the scatter $\sigma_N$ as a function of the noise of the unbinned data $\sigma_1$ is
\begin{equation}
  \label{eq:uncorrelated_data}
  \sigma_N = \beta \frac{\sigma_1}{\sqrt{N}} \sqrt{\frac{M}{M-1}}\, .
\end{equation}
For uncorrelated data, $\beta = 1$. Our residuals have an average of $\beta=1.15$, suggesting minimal time correlated trends in the residuals. Figure~\ref{fig:residual_bin} shows the RMS of the residuals as a function of bin widths. The lack of need for any decorrelation can be primarily attributed to the use of variable, elliptical, apertures. 

\section{VSTAR atmosphere model}
\label{sec:vstar-model}

We use the VSTAR line-by-line radiative transfer code \citep{2012MNRAS.419.1913B} to derive a model atmosphere of WASP-19b that fits our measurement and the data published previously. VSTAR is a comprehensive atmospheric radiative transfer model incorporating a chemical equilibrium model, an extensive database of molecular spectral lines and a full treatment of multiple-scattering radiative transfer using the discrete ordinate method. It has been extensively tested and applied to objects ranging from solar system planets \citep{2013Icar..222..364C,2012Icar..217..570C,2011MNRAS.414.1483K} to M-dwarfs \citep{2012MNRAS.419.1913B}. It is impossible to obtain a unique model that can best fit the currently available broadband data that only sparsely covers the optical and infrared spectrum. Instead we focus on a discussion of effects observed in a spectrum by changing specific conditions in the planetary atmosphere. 

Highly irradiated planets, like WASP-19b, have been hypothesised to show thermal inversion in its atmospheric profile due to condensation of VO and TiO within a cold trap \citep{2008ApJ...678.1419F}. However the Spitzer IRAC data \citep{2011arXiv1112.5145A} and near infrared ground measurements at 1.6 and 2.1 $\mu$m \citep{2010A&amp;A...513L...3A,2010MNRAS.404L.114G} appear to be inconsistent with thermal inversion in the planet's atmosphere. Various explanations have been proposed for the lack of thermal inversion in some highly irradiated planets, such as the dependency on the presence of a cold trap \citep{2009ApJ...699..564S,2009ApJ...699.1487S}, destruction of absorbers by stellar activity \citep{2010ApJ...720.1569K}, disequilibrium photochemistry \citep{2009ApJ...701L..20Z}, or the enrichment of C/O that leads to depleted TiO abundance \citep{2012ApJ...758...36M}.

\begin{figure*}[h]
  \centering
  \includegraphics[width=8cm]{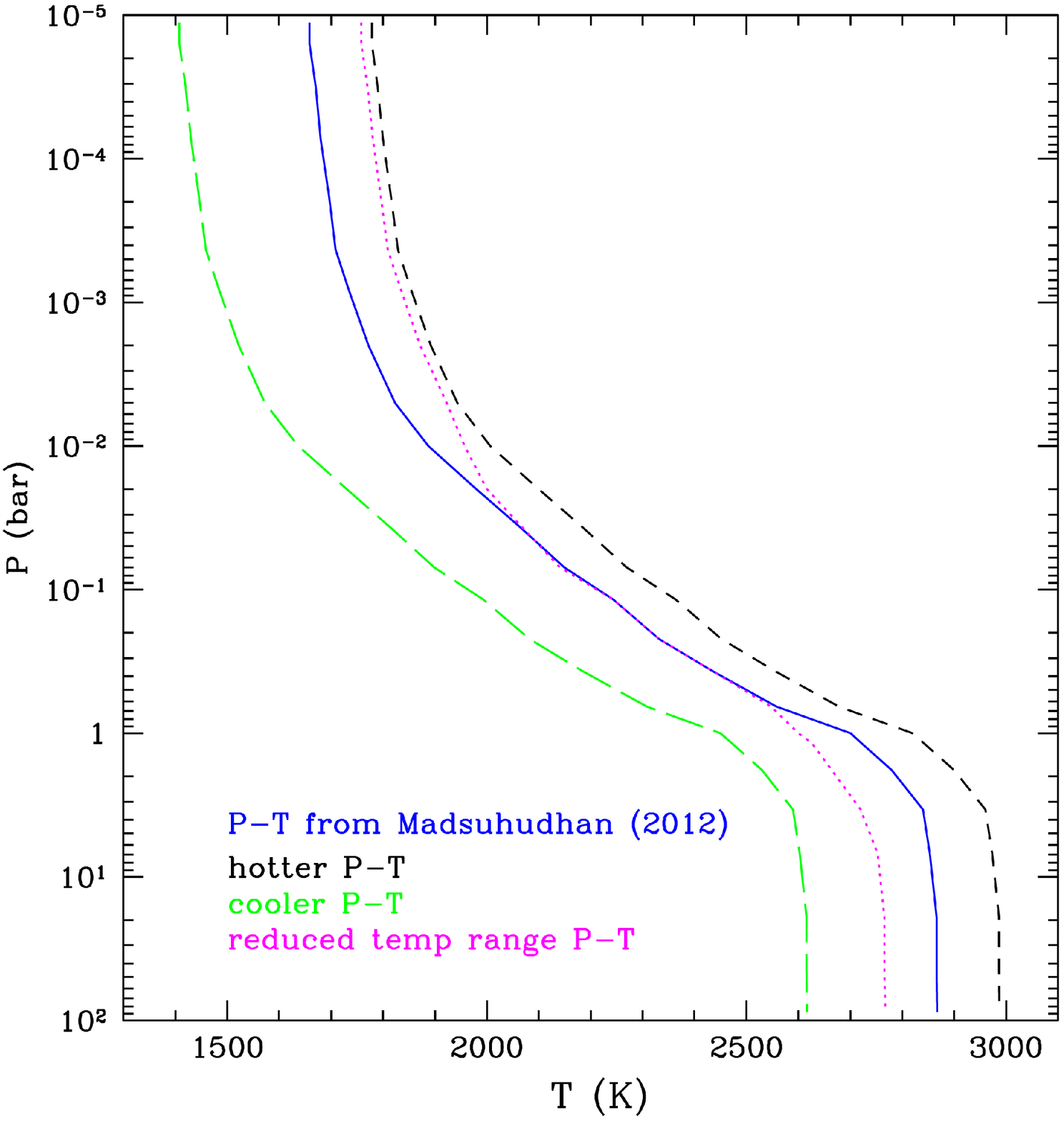}
  \includegraphics[width=8cm]{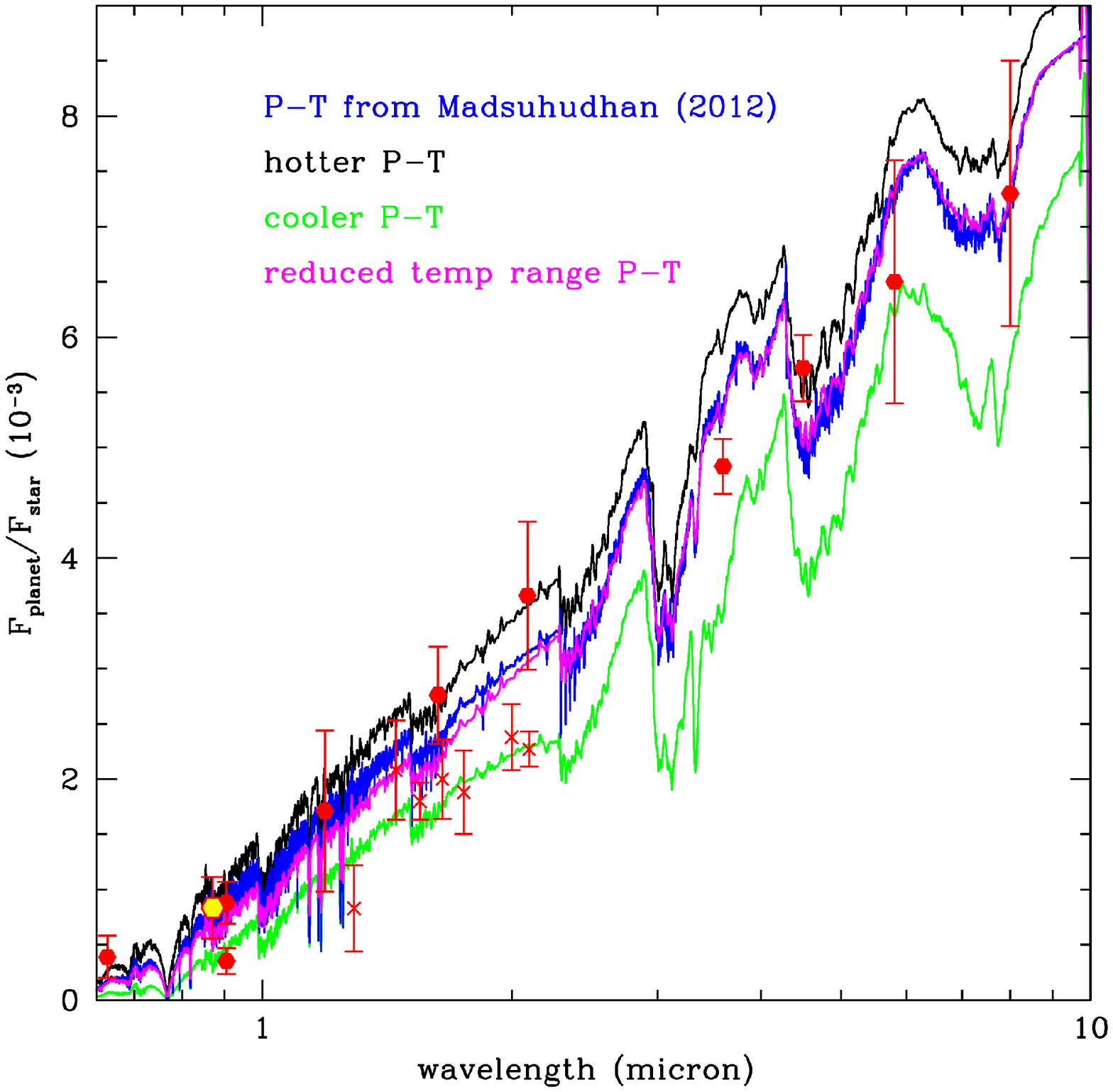}
\caption{Left: Four P-T atmosphere profiles described in more detail in Section~\ref{sec:vstar-model}. Right: The WASP-19b VSTAR models corresponding to the P-T profiles shown on the left. C/O ratio of 1.1 was assumed in all four models and molecular absorption due to H$_{2}$O, CO, CH$_{4}$, CO$_{2}$, C$_{2}$H$_{2}$, HCN, TiO, VO. The yellow hexigon denotes the data point from the FTS observation reported in this paper, red crosses show data from \citet{2013arXiv1303.1094B}, while red hexagons mark results of all other observations described in Section~\ref{sec:introduction}. }
  \label{fig:atmos}
\end{figure*}

In our modelling we use four different atmospheric pressure--temperature (P-T) profiles without inversion (Figure~\ref{fig:atmos}a): (I) the red profile in Fig.12 of \citet{2012ApJ...758...36M}; (II) a `hotter' profile, which corresponds to conditions discussed by \citet{2011arXiv1112.5145A}; (III) a `cooler' profile which reflects range of temperatures and pressures assumed by \citet{2013arXiv1303.1094B} to explain their near infrared data; (IV) a `narrow' P-T profile, with reduced range of temperatures that overlap with the \citet{2012ApJ...758...36M} model over the range of 0.05-0.5 bar.

Our models assume a plane-parallel, stratified atmosphere with 25 layers characterised by temperature, pressure and mixing ratios of the following molecular and atomic species: H$_{2}$O, CO, CH$_{4}$, CO$_{2}$, C$_{2}$H$_{2}$, HCN, TiO, VO, Na, K, H$_{2}$, He, Rb, Cs, CaH, CrH, MgH and FeH.  Mixing ratios of these opacity sources are calculated in chemical equilibrium.  Atmospheres of hot-Jupiters like WASP-19b are most likely dominated by H$_{2}$, which is a source of the H$_{2}$-H$_{2}$ and H$_{2}$-He collisionally induced absorption (CIA) that we included with opacities calculated by \citet{1998STIN...9963017B}. We also considered Rayleigh scattering by H$_{2}$, He, H in the atmosphere of the planet, and free-free and bound-free absorption from H, H$^{-}$ and H$_{2}$$^{-}$. Our spectral line absorption database is described in detail in \citet{2012MNRAS.419.1913B}. Table~\ref{tab:refsline} lists the references to the sources of spectral lines for absorbers used in our models. A spectrum of the WASP-19, G8V type star was obtained from the STScI stellar atmosphere models by \citet{2004astro.ph..5087C}.

\begin{deluxetable}{lp{5cm}}
\tablewidth{0pc}
\tabletypesize{\scriptsize}
\tablecaption{
    List of molecular and atomic absorbers used in the VSTAR modelling with references to the line databases.
    \label{tab:refsline}
}

\tablehead{
    \multicolumn{1}{c}{Line}          &
    \multicolumn{1}{c}{Reference }             \\
    \multicolumn{1}{c}{Absorbers}          &
    \multicolumn{1}{c}{}             \\
}
\startdata
    CH$_{4}$ & see 2.2.6 in \citet{2012MNRAS.419.1913B}\\
    CO$_{2}$ & \citet{1993Icar..103....1P} \\
    H$_2$O & \citet{2006MNRAS.368.1087B} \\
    CO & \citet{1994ApJS...95..535G} \\
    HCN &\citet{2006MNRAS.367..400H} \\
    C$_{2}$H$_{2}$ & \citet{2009JQSRT.110..533R}\\
    CaH& \citet{2003ApJ...582.1263W} \\
    MgH & \citet{2003ApJ...582.1059W,2003ApJS..148..599S} \\
    FeH & \citet{2003ApJ...594..651D,2010AJ....140..919H} \\
    CrH &\citet{2002ApJ...577..986B} \\
    TiO& \citet{1998Aamp;A...337..495P}\\
    VO& Plez, B., private communication\\
    K, Na, Rb, Cs & \citet{1995Aamp;AS..112..525P,1999Aamp;AS..138..119K} \\
\enddata 

\end{deluxetable}

Figure~\ref{fig:atmos}b shows the model spectra for the four P-T profiles considered above. All models presented in this figure have  carbon to oxygen ratio C/O=1.1. Thus they can be easily compared with the red spectrum in Fig.12 of \citet{2012ApJ...758...36M}. The differences between our model and the \citet{2012ApJ...758...36M} model with the same P-T profile can be attributed to use of line databases which may have varied levels of completeness. The model with `hotter' P-T profile tends to fit better near infrared (NIR) data from \citet{2010A&amp;A...513L...3A} and \citet{2010MNRAS.404L.114G}, while `cooler' P-T profile produces a spectrum that matches closer to the data obtained by \citet{2013arXiv1303.1094B}. However both these profiles either overestimate or underestimate absorption observed in the Spitzer data between 3.6 and 8 $\mu$m. We also found very little difference between the `narrow' P-T profile and the one from \citet{2012ApJ...758...36M}, with only slightly increased absorption between 1 and 2$\mu$m in the `narrow' P-T profile.

\begin{figure*}[h]
  \centering
   \includegraphics[width=8cm]{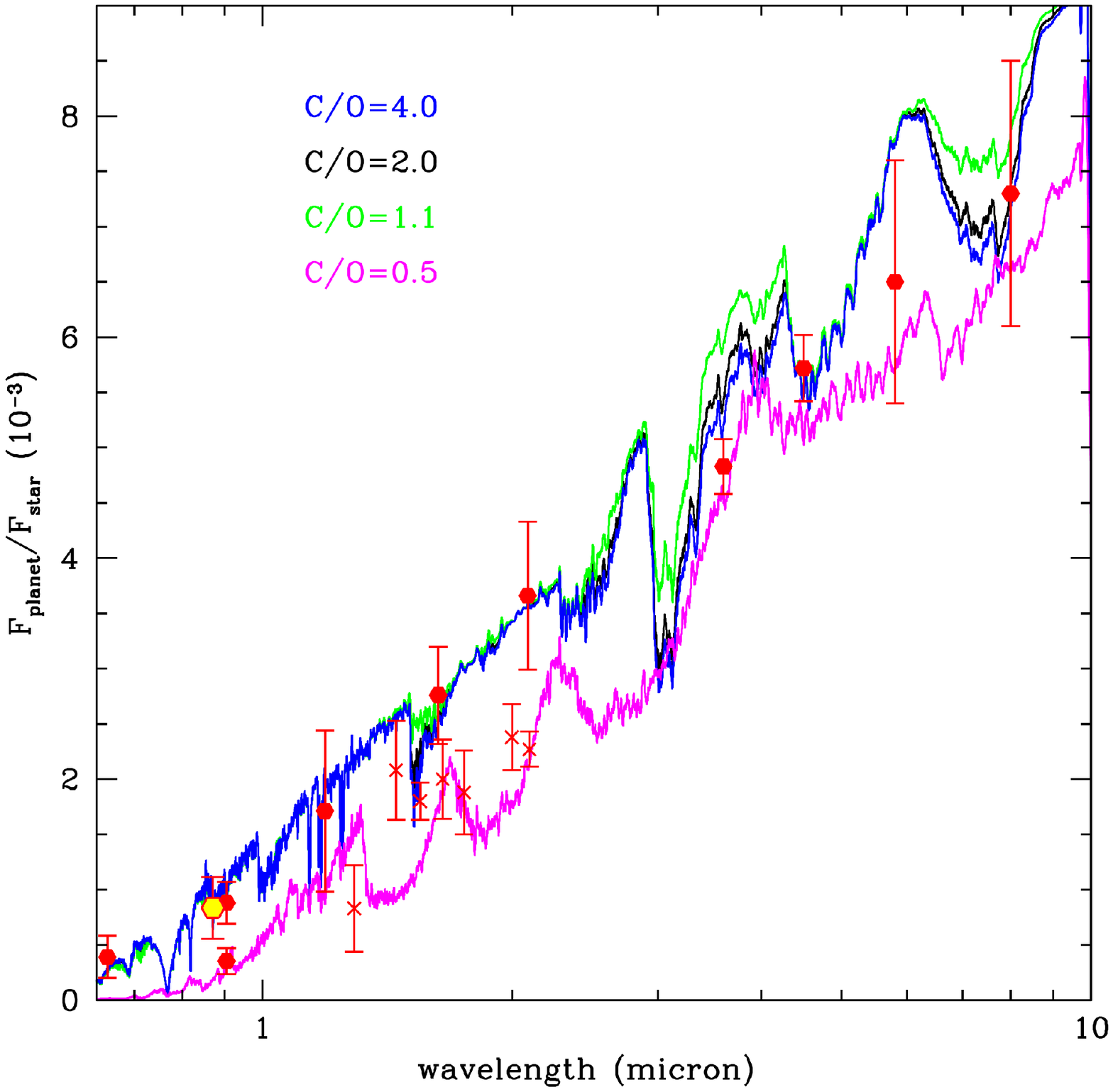}
    \includegraphics[width=8cm]{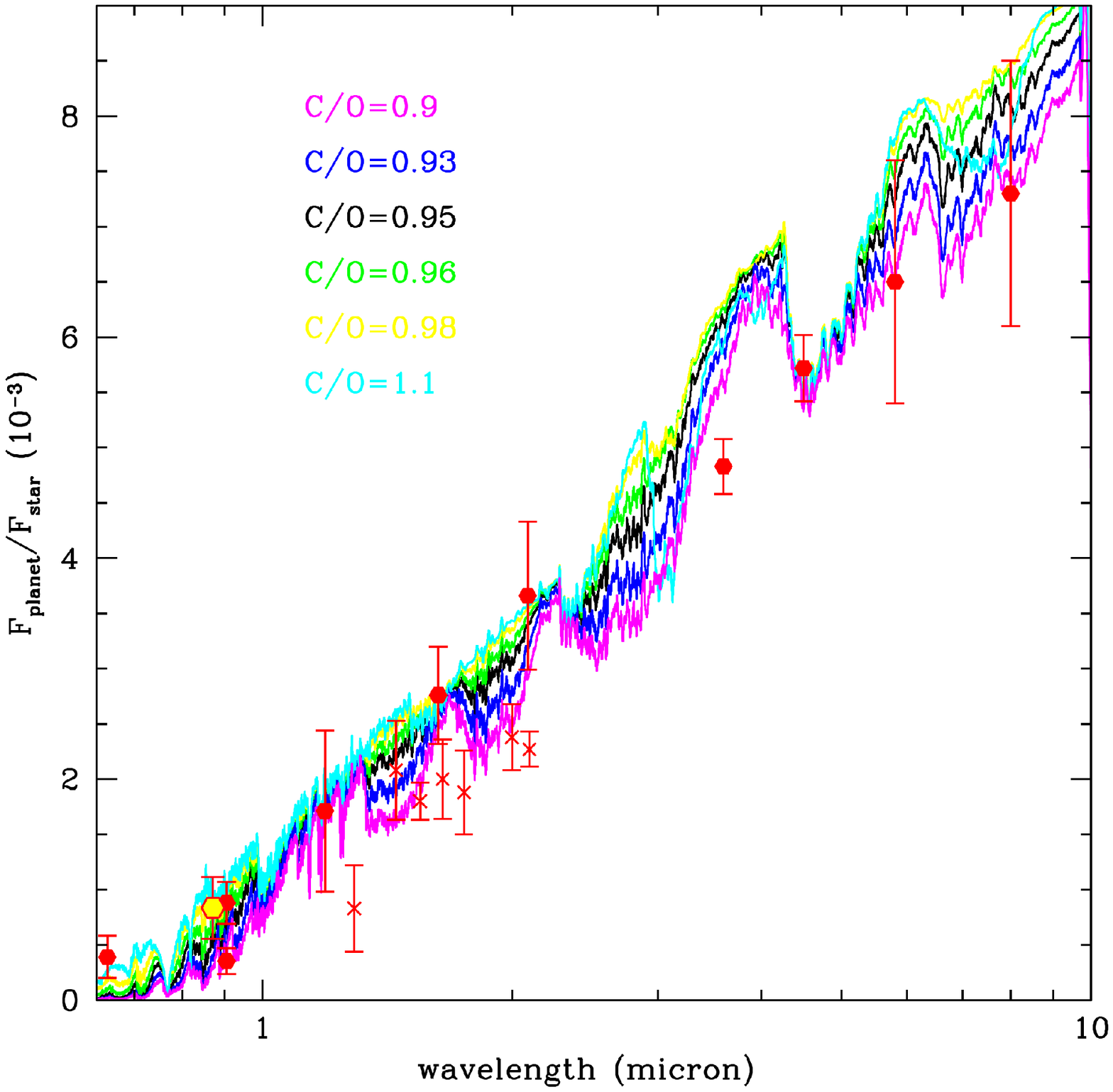}
  \caption{Example of WASP-19b atmosphere models with for the `hotter' P-T profile and changing ratio of C/O. Left: The model with close to solar ratio C/O=0.5 seems to be especially inconsistent with the data from optical observations. Models with C/O ratios above 1 fit better all data except the NIR observations of \citet{2013arXiv1303.1094B}. Right: Strong changes in water absorption bands across infrared range of the spectrum around the region of C/O=1 are shown, where abundances of oxygen and carbon bearing species vary by orders of magnitude.}
  \label{fig:atmos1}
\end{figure*}

In Figure~\ref{fig:atmos1} we show the effect of varying C/O ratio on spectra using example of a model with our hottest P-T profile, although qualitative results are the same for other profiles considered here. The spectrum obtained for C/O=0.5 is dominated by the oxygen bearing molecules with strong H$_{2}$O bands visible in NIR and far IR,  and CO$_{2}$ and CO bands around 4.5 $\mu$m. Around  C/O=1 the abundances of carbon and oxygen containing molecules change dramatically by many orders of magnitudes. This explains the rapid decrease of H$_{2}$O absorptions in the spectra when C/O ratio varies between 0.9 and 1.1  in right panel of Figure~\ref{fig:atmos1}, while only modest changes are visible in left panel of Figure~\ref{fig:atmos1} between C/O=1.1 and C/O=4.0. Spectra of atmospheres with high content of carbon are dominated by CH$_{4}$ absorption in addition of molecules such as HCN and C$_{2}$H$_{2}$ considered also in \citet{2012ApJ...758...36M}. While strong water absorption bands are absent, the CO features around 2.3 and 4.8 $\mu$m become more prominent. Currently available spectral measurements for WASP-19b seem to be more consistent with the atmosphere models which are derived with the C/O ratio higher than solar. 

Models presented so far in Figure~\ref{fig:atmos} assumed a clear atmosphere. However recently published data for the hot-Jupiter HD189733b \citep{2008MNRAS.385..109P} indicate a presence of haze in the top layers of its atmosphere. Composition of hazes depends on the abundance and refractory properties of different compounds \citep{1999ApJ...512..843B}.  In hot-Jupiters and brown dwarfs suggested condensates may be formed by highly refractory species such as perovskite (CaTiO$_{3}$) and corundum (Al$_{2}$O$_{3}$), which condense in temperatures close to 1600~K. More abundant Si, Mg and Fe elements combine into compounds such as enstatite (MgSiO$_{3}$) and forsterite (Mg$_{2}$SiO$_{4}$) that condense in lower temperatures.  
Even at the relatively lower temperatures at the top of the atmosphere of WASP-19b it is not clear which species could potentially exists in a form of haze. 

\begin{figure*}[h]
  \centering
  \includegraphics[width=8cm]{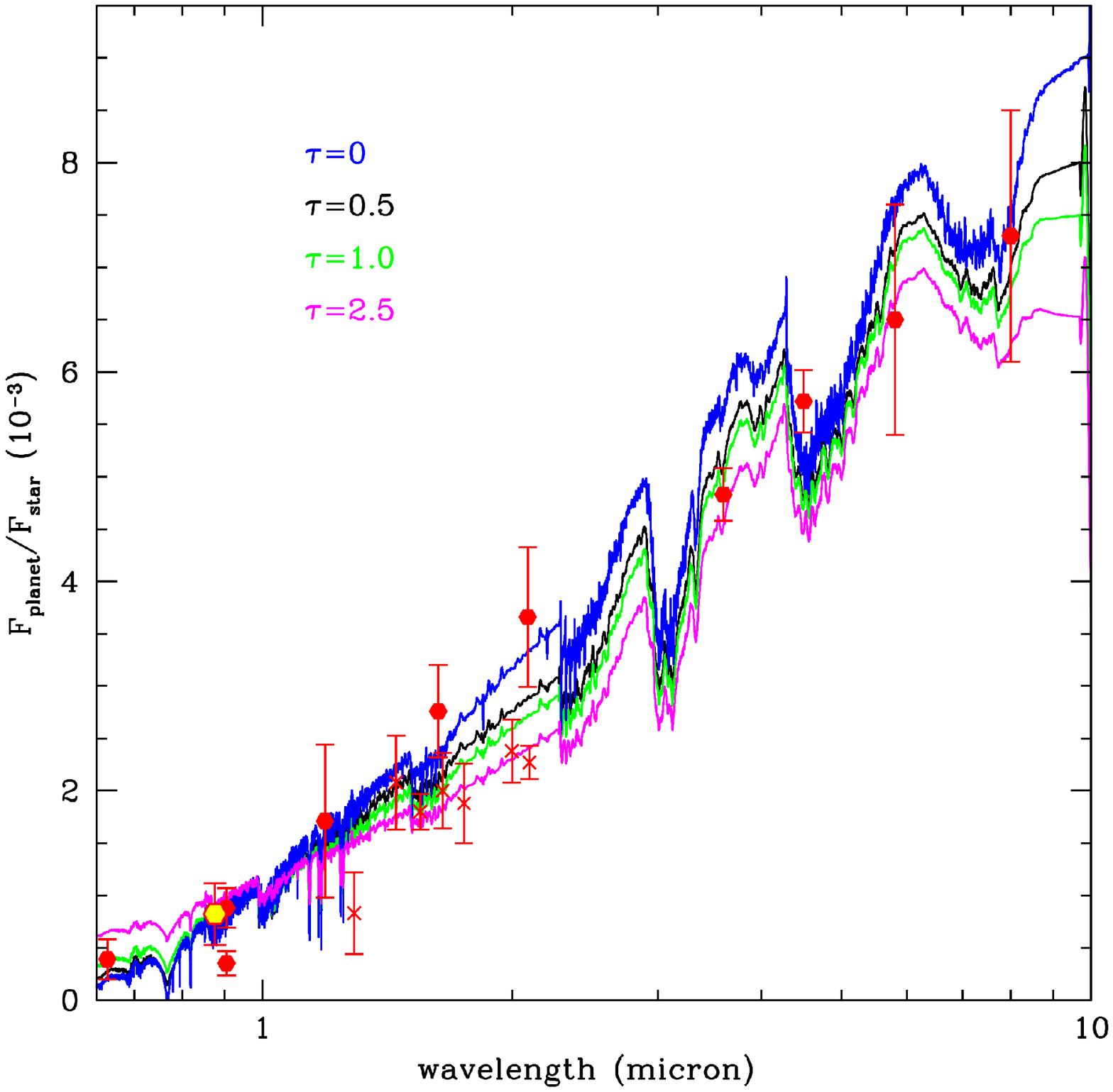}
   \includegraphics[width=8cm]{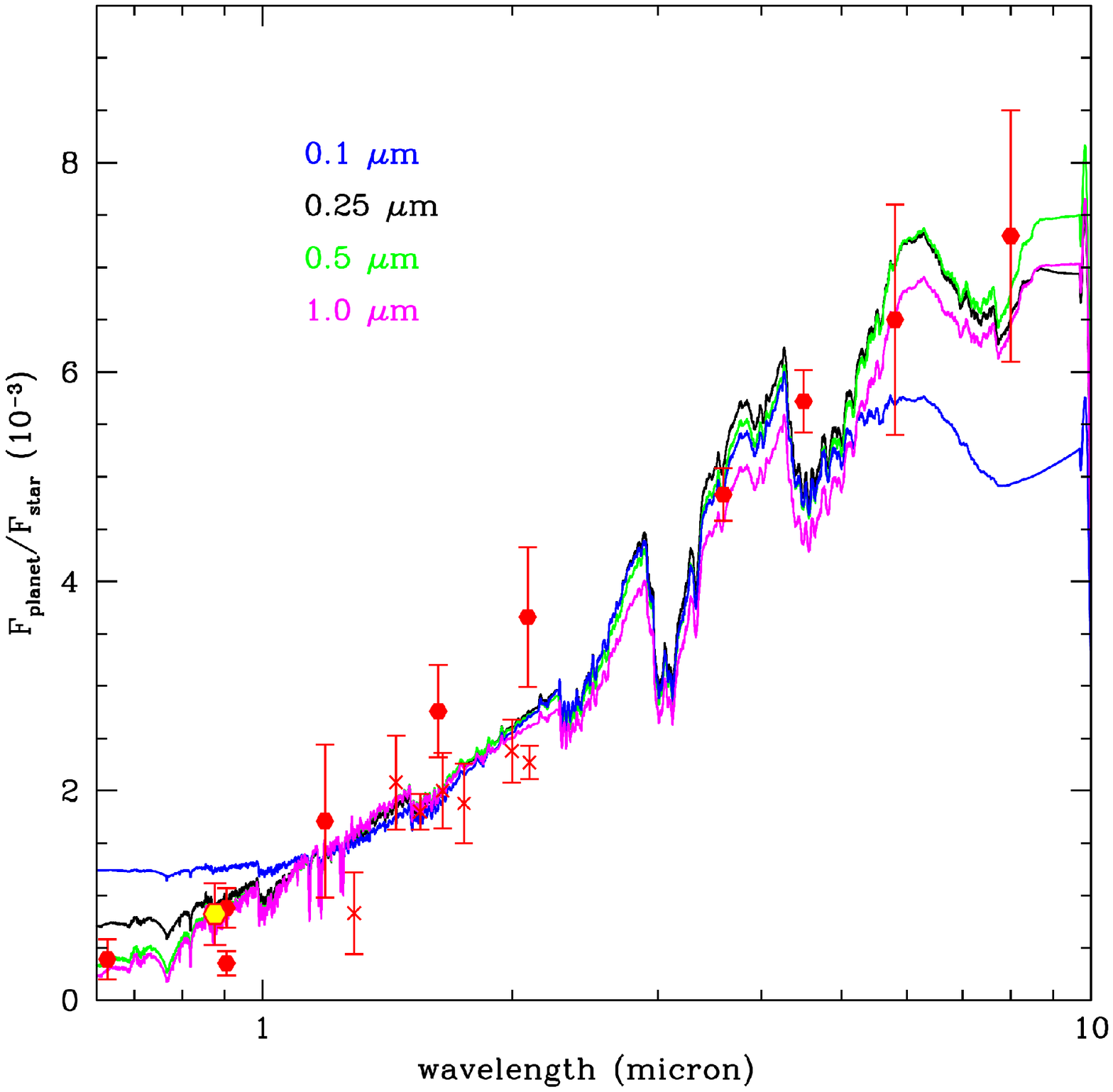}
  \caption{Model of the WASP-19b atmosphere with P-T profile from  \citet{2012ApJ...758...36M} shown in Figure~\ref{fig:atmos} and C/O=1.1. Left: Including clouds of varied opacity ($\tau$) at the top of atmospheric layer, with a power-law distribution of particles with effective radius of 0.5 $\mu$m and effective variance 0.2 $\mu$m as defined in \citet{2002sael.book.....M}. Right: Varying the mean particle size, whilst assuming $\tau=1$ at $1.5\,\mu\text{m}$.}
  \label{fig:tau}
\end{figure*}

In Figure~\ref{fig:tau} we assume that such a haze exists and it is composed of unknown particulate with refractive index similar to enstatite. Four examples of the model spectrum are shown for WASP-19b, where the optical depth of a cloud in the top layer of the atmosphere is varied. The particles with a mean size of 0.5 $\mu$m are assumed in the left panel. On the right all models are derived with the varied mean size of particles, while the same optical depth $\tau=1$ at 1.5 $\mu$m is assumed. The absorption and scattering properties as a function of wavelength are calculated using Mie theory. At wavelengths comparable to the size of cloud particles,  scattering processes operate efficiently, which leads to increase of planetary albedo in the corresponding part of its spectrum.  On the other hand the added opacity in top layers obscures thermal emission from the planet, which has an effect of lowering the received flux in infrared part of a spectrum. 
Differences in particle sizes affect both scattering and absorption properties of the haze. Particles smaller than  0.5 $\mu$m appear to generate highly reflective haze at visible wavelengths, which may not be consistent with the measurement from \citet{2013arXiv1303.0973A}. 

Observations of secondary eclipses at different wavelengths are sensitive to different properties of the planetary atmosphere. Observations of the flux in z-band can provide a sensitive probe of the C/O ratio in the atmosphere of the planet, as shown in Figure~\ref{fig:atmos1}. In optical spectrum strongly absorbing bands of VO and TiO dominate the measured flux in temperatures above 1700~K  when the C/O ratio is similar to solar. After VO and TiO start to condense below this temperature, absorption from alkali lines and water bands takes over. However if C/O is higher than solar, alkali lines will be distinct even at lower temperatures due to reduced abundance of VO and TiO.   
On the other hand, in the presence of stratospheric haze Rayleigh scattering may dominate optical spectrum almost entirely as seen in Figure~\ref{fig:tau}. 
A few more strategically placed photometric data points in optical and near infrared spectrum will help to discriminate between these  broad conditions of the WASP-19b atmosphere. However more detailed and unique models can only be derived when the amount of photometric data becomes sufficient to break degeneracies in interpretation of current spectral features. This is currently rather remote prospect as discussed in \citet{2013arXiv1304.5561L}. 

\section{Discussions}
\label{sec:discussions}

We presented an examination of the emission spectrum of WASP-19b measured in eclipse. Using FTS observations, we measured the $z$ band eclipse depth to be $0.080\pm0.029\,\text{\%}$. This result is in excellent agreement with the depth measured by \citet{2012ApJS..201...36B} of $0.088\pm0.019\,\text{\%}$, and also consistent with the tentative detection of a significant eclipse in the optical ASTEP band by \citet{2013arXiv1303.0973A}, as well as deep NIR detections by \citet{2010A&amp;A...513L...3A,2010MNRAS.404L.114G}. It is also in $2\sigma$ agreement with the measurement made using multiple eclipses from the 1.2\,m Euler-Swiss telescope and the 0.6\,m TRAPPIST telescope of $0.035\pm0.012\,\text{\%}$ \citep{2012arXiv1212.3553L}. 

From the non-exhaustive set of VSTAR spectra, we find no single model that can fit all of the reported observations. However, when the spectrophotometry measurements by \citet{2013arXiv1303.1094B} are discarded, the C/O enriched models present a good fit to the remaining points. The \citet{2013arXiv1303.1094B} points are also inconsistent with available photometric $J,H,K$ measurements at the same wavelengths. We also note that our new $z$ band detection is consistent in brightness temperature with the photometric near-infrared detections, not the spectrophotometry measurement. These difficulties highlight the challenges of transit spectrophotometry observations, especially when WASP-19 is the faintest object targeted with the technique to date. 

\citet{2013arXiv1304.5561L} assessed the difficulty of interpreting broadband emission spectra via retrieval techniques, and noted that C/O classifications tend towards a bimodal posterior of 0.5 or 1. This agrees with our assessment that large-scale changes in the spectrum are only apparent from C/O of 0.9 to 1.1. Although a quantitative estimate of carbon enrichment in these atmospheres is unlikely, WASP-19b is still more consistent with a super-solar C/O composition. 

In addition to WASP-19b, \citet{2012ApJ...758...36M} pointed to XO-1b, CoRoT-2b, WASP-33b, and WASP-12b as carbon rich candidates. XO-1b is a significantly less irradiated planet that has only been studied in the Spitzer bands \citep{2008ApJ...684.1427M}. WASP-33 is a rapidly rotating F-dwarf that exhibits photometric variability on the hour timescale, for which precision photometry results are difficult to interpret \citep{2011MNRAS.416.2096S}. Light from WASP-12 was found to be contaminated by a blended M-dwarf, and the compensated eclipse measurements can be modelled without a carbon rich atmosphere \citep{2012ApJ...760..140C}. Only CoRoT-2b has received as thorough an observational evaluation as WASP-19b, with measurements available from the CoRoT optical band to the Spitzer bands. No existing analysis has included all available observations to examine the validity of its carbon enriched claim.

\acknowledgments
This paper uses observations obtained with facilities of the Las Cumbres Observatory Global Telescope. The work was, in part, supported by the Australian Research Council through Discovery grant DP110103167.

Facilities: \facility{FTS(Merope)}

\bibliographystyle{apj}

\end{document}